\documentstyle[11pt]{article}
\begin{document}
\oddsidemargin 0.2 in
\evensidemargin 0.2 in
\newcommand{\1}{\begin{equation}}
\newcommand{\2}{\end{equation}}
\newcommand{\3}{\begin{eqnarray}}
\newcommand{\4}{\end{eqnarray}}
\newcommand{\è}{\`{e}}
\newcommand{\é}{\'{e}}
\newcommand{\à}{\`{a}}
\newcommand{\ò}{\`{o}}
\newcommand{\ù}{\`{u}}
\baselineskip=21pt
\title{Jet dynamics in black hole physics: 
acceleration during subparsec collimation\footnote{Whork partially supported by Ministero della Ricerca Scientifica e Tecnologica of Italy and by Gruppo Nazionale della Fisica Matematica del CNR}}
\author{Fernando de Felice 
\footnote{Department of Physics G. Galilei, University of Padova,
via Marzolo 8, I-35131 Padova Italy and
INFN Sezione di Padova. 
E-mail: defelice@ pd.infn.it} and Olindo Zanotti\footnote{
International School for Advanced Studies, SISSA, Trieste, Italy.
E-mail: zanotti@sissa.it}}

\maketitle

\begin{abstract} 
We study the processes of particle acceleration which take place 
in the field of a
rotating black hole as part of a mechanism of formation  of galactic jets   
within the first parsec from the central source, where gravitation is 
supposed to be dominant. 
  We find the 
Lorentz factor that a stream of particles acquires as function of distance, 
when the orbital parameters vary slightly due to a local electromagnetic 
field or a pressure gradient.
\end{abstract}
\section{Introduction}
Jets emerging from active galactic nuclei are highly collimated structures 
(as revealed by  radio maps) probably made of electron-positron plasma, 
which propagate in the intergalactic medium with relativistic velocities 
along most of their length (as implied by the detection of superluminal motion). 
It is then clear that one has to search for mechanisms which allow 
for both collimation and acceleration.

On large scales (typically from parsec to kiloparsec) jets are 
currently studied within the framework of magnetohydrodynamics, 
to which we owe most of our knowledge about these structures. 
Through analytical and numerical models, we are now able to say something 
about 
electrodynamic confinement  of axisymmetric flows, 
electromagnetic extraction of energy 
from rotating black holes through force-free magnetospheres and asymptotic 
poloidal velocities of hydromagnetic flows, to mention but a few. What is 
shared by most of these models is the need for plasma injection into a 
rapidly rotating magnetosphere from below. 

Attention is then shifted to the subparsec scale, just outside the event 
horizon, where a primary acceleration mechanism must be  at work.
Recent developments (Sikora et al. 1996, \cite{a}) have shown that such 
a mechanism cannot be powered by the radiation pressure of the accretion 
disk; this pressure, instead, causes the flow in the jet to decelerate by 
virtue of 
inverse Compton scattering with the plasma of the jet itself, with a 
maximum efficiency
 when the plasma's Lorentz factor has reached values higher than an 
 equilibrium $\gamma_{eq}\sim 4$. That is to say, jets, accelerated in the 
subparsec region up to $\gamma_j\geq 5$, as revealed by VLBI measurements of 
superluminal motion in extragalactic radio sources,  cannot avoid radiation 
drag. Notwithstanding this, 
we expect  that at subparsec scale the gravitational field of a 
$10^{8}\div10^{11}M_{\odot}$ black hole should still play a major role in 
determining  particles' motion; indeed the behaviour of individual 
particles is also that of the bulk of fluid elements in the guiding centre 
approximation. 
We then  studied the combined 
effects of gravity and external physical perturbations.
In  de Felice and Carlotto (1997, \cite{b}; hereafter Paper I) the 
collimating behaviour of geodesic orbits in the presence of a constrained 
variation of their energy and angular momentum was considered. The constraints 
are those 
which allow a particle, initially on a geodesic, to move on a nearby 
geodesic characterized by {\it slightly} varied parameters. In Paper I, 
this requirement was termed {\it geodesicity condition}.
Following this line of thought, Karas and Dovciak (1997, \cite{c}) have 
estimated the rate of change of the
orbital parameters of individual particles and have integrated this rate 
over a power law distribution of particles' energy. Their results confirm our claim 
in  Paper I that the approximation of geodesic motion in the presence of 
small perturbations, is 
appropriate for modelling the primary collimation of a jet.

Here we shall investigate whether the geodesicity conditions considered in Paper I, are compatible with the local Lorentz factor which is observed in galactic jets. Indeed we show that, provided there is a fine 
tuning between the stiffness of geodesic orbits and the effects of an 
external field, a large family of particle trajectories described by the 
Kerr metric not only collimate towards the axis of symmetry (see Paper I), 
but also accelerate, reaching values of the Lorentz factor $\gamma$, 
as measured by a local static observer, which are consistent with 
observations. Typically, we find $\gamma\leq 10$ at $1pc$ from the centre.

In Section 2 we summarize the general relativistic collimation effect 
discussed in Paper I,
then in Section 3 we analyse the acceleration which test particles  acquire under the condition of the 
mentioned collimation process. Behaviours of the local Lorentz factor 
$\gamma$ are found as functions of distance from the central source for both 
the cases of 
Lorentz forces arising from a local electromagnetic field and from pressure 
gradients. Comparison with observations must be handled with care: some 
comments on this problem are made in Section 4. Finally in Section 5, we pay 
particular attention to the limits imposed by the geodesicity conditions on the distance scale where the collimation mechanism is allowed to operate.

\section{Geometrically induced collimation}
In Paper I, de Felice and Carlotto studied the tendency of vortical 
geodesics, in a Kerr background geometry, to collimate towards the axis of 
symmetry under a constrained variation of the constants of the motion. 
Their claim was that such a property might be astrophysically relevant to 
allow for jets primary collimation very close to the central black hole. 
Here we shall summarize that reasoning.

Let the space-time be described 
by the Kerr metric in Boyer and Lindquist coordinates $(t,r,\theta,\varphi)$:
\begin{eqnarray}
ds^{2}&=&-(1-\frac{2Mr}{\Sigma})dt^{2}-\frac{2\cal A}{\Sigma} 
\omega \sin^{2}{\theta}\,dtd\varphi+\nonumber \\ && +\frac{\Sigma}
{\Delta}dr^{2} +\Sigma d\theta^{2}+\frac{\cal A}{\Sigma}\sin^{2}{\theta}
\,d\varphi^{2} \label{kerr}
\end{eqnarray}
where
\begin{eqnarray}
\Sigma&=&r^{2}+a^{2}\cos^{2}{\theta}\\
\Delta&=&r^{2}+a^{2}-2Mr\\
\cal A&=&(r^{2}+a^{2})^{2}-a^{2}\Delta\sin^{2}{\theta}\\
\omega&=&\frac{2Mar}{\cal A}
\end{eqnarray}
$M$ and $a$ being respectively the total mass energy of the metric source 
and its 
specific angular momentum ($a=J/cM$), both expressed in geometrized 
units ($c=G=1$).

Attention is focused on a particular family of geodesics, namely  
the vortical ones. These curves are gravitationally unbound 
(open orbits) and are characterized by the following conditions
\footnote{Vortical geodesics are also those with
$\Gamma=l=L=0$, but we shall not consider them here.}:
\1
\Gamma>0\hspace{1cm}-a^{2}\Gamma\leq L\leq a^{2}\Gamma\hspace{1cm}
L<l^{2}\leq\frac{(L+a^{2}\Gamma)^{2}}{4a^{2}\Gamma}\label{cond}
\2
In the absence of any external perturbation, $l$ and $E=\sqrt{\Gamma+1}$ 
are constants of the motion and express, respectively, the azimuthal angular 
momentum (in units of $\mu c$) and the total energy (in units of $\mu c^{2}$)
of the particle along the orbit. $L$ is the separation constant of the 
Hamilton-Jacobi equation in the Kerr metric and is related to the square 
of the total angular momentum of the particle (de Felice, 1980, \cite{d}; 
de Felice and Preti 1998, \cite{h}).

\begin{figure}
\begin{center}
\input{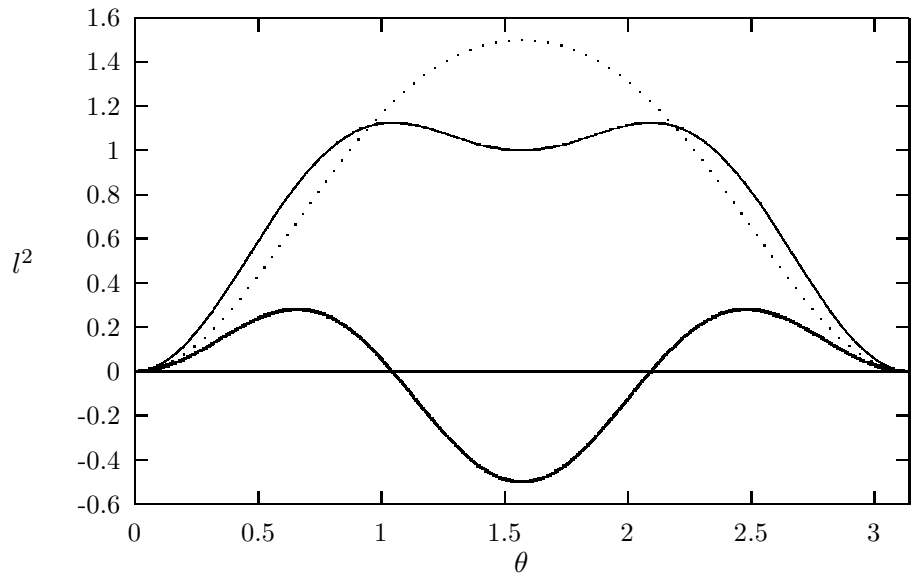}
\caption{Plots of the functions $l^{2}=l^{2}(\theta,\Gamma,L)$, 
when $\Gamma>0$.}\label{plots}
\end{center}
\end{figure}

The permitted values of $\theta$  for the geodesic motion are confined (see figure \ref{plots}) to the  area below the functional curves:
\1
l^{2}(\theta,L,\Gamma)=\sin^{2}{\theta}(L+\Gamma a^{2}\cos^{2}{\theta})
\label{curv}
\2
in the $(l^{2},\theta)$ space (for any fixed pair of $L$ and $\Gamma$), 
along which $\dot\theta=0$.
Given a constant value of $l^{2}$ in accordance with condition (\ref{cond}), 
vortical motion is found to be latitudinally confined within the range 
$[\theta_{1},\theta_{2}]$ determined by the intersection of the straight 
line $l^{2}=const$ with the functional curves (\ref{curv}).
That is the reason why, vortical geodesics, which never cross the equatorial 
plane, are the most likely to leave the innermost part of an accretion disk 
surrounding a rotating black hole, through a spiralling motion.

Collimation is studied with respect to the opening angle of a particle beam 
centred on the axis $(\theta=0)$, taken for convenience to be equal to the 
angle which makes  $l^{2}$  vanish:
\1
\cos^{2}{\theta_{0}}=-\frac{L}{\Gamma a^{2}}.\label{zeri}
\2

The cornerstone of the reasoning is to presume that, in the innermost 
region of the field, relation (\ref{zeri}) continues to hold even under a 
slow variation of the parameters $\Gamma$ and $L$ induced by some sort 
of external perturbation to the plain geodesic motion.

In order to prevent  the geodesic character of the motion being 
lost because of the perturbations, we will impose 
``geodesicity conditions" with the effect of forcing each 
particle on a vortical geodesic to drift onto a nearby geodesic of the same 
type.

In this case, differentiation of (\ref{zeri}) with respect to the 
proper time $\tau$ along the trajectories, leads to the basic equation of 
jet dynamics:
\1
- 2L\tan{\theta_{0}}\frac{d\theta_{0}}{d\tau}=
a^{2}\cos^{2}{\theta_{0}}\frac{d\Gamma}{d\tau}+\frac{dL}{d\tau}.
\2

The existence of vortical trajectories in the space-time of a Kerr black hole,
 is a general relativistic effect which stems from the rotational properties 
of the metric, (see also O'Neil (1995) \cite{o}). While the vortical 
character of the orbits with parameters as in (6), is a natural consequence 
of  gravitational dragging, the trend to axial collimation, as a result of a 
small perturbation, was quite unexpected. This effect is entirely due to the 
first term on the right-hand-side of equation (9) which contains the 
rotational parameter $a$. Evidently, when $a=0$, there are no vortical orbits 
and no axial collimation; this implies that sufficiently small values of $a$ 
would make the effect
negligibly small. However, if we consider, as source of the orbital 
perturbations, physical fields that share the same symmetries as  the 
metric source, as expected very near to the black hole, the collimation 
follows laws which do not depend explicitely on $a$ (see equations (10) to 
(12) below), suggesting a kind of contradiction. Indeed, the dependence on 
$a$ comes implicitly through the very form of the equations of motion and 
the equations of the perturbing fields. Parameter $a$ fixes, 
from relations (6), the range of the permitted vortical orbits; when $a$ 
becomes small, this range shrinks (see figure 1), so what decreases is not 
the {\it amount } of collimation, but the {\it number} of orbits which are 
involved.  For this reason, this general relativistic effect may  not be 
negligible at astrophysical scales.

\begin{figure}
\begin{center}
\input{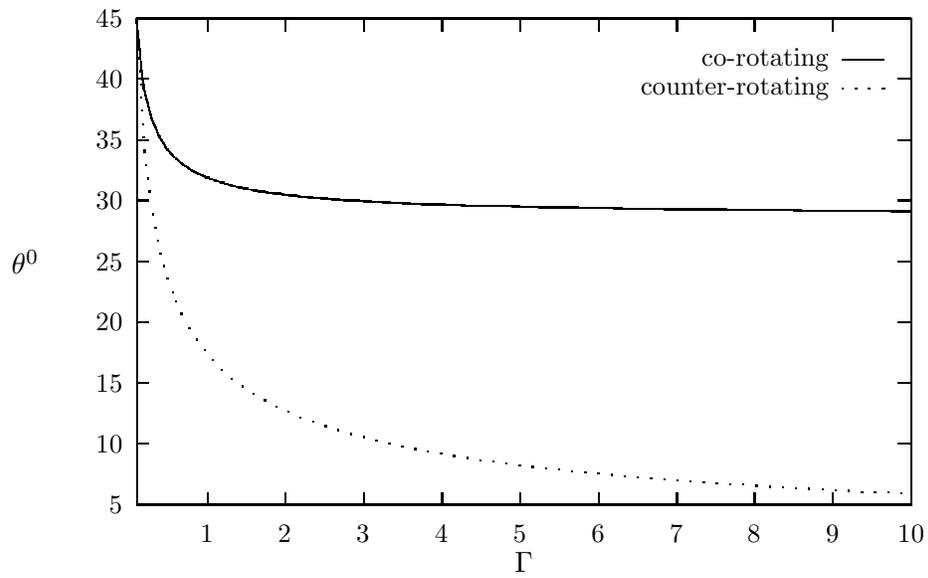}
\caption{\small{Behaviour of the angle of collimation  
for corotating  and counter-rotating  
outgoing particles which increase their energy with respect to infinity, 
under the influence of a poloidal magnetic field. 
$\Gamma_{i}=0.1$, $\theta_{i}=\pi/4$.}}\label{coll1}
\end{center}
\end{figure}

\begin{figure}
\begin{center}
\input{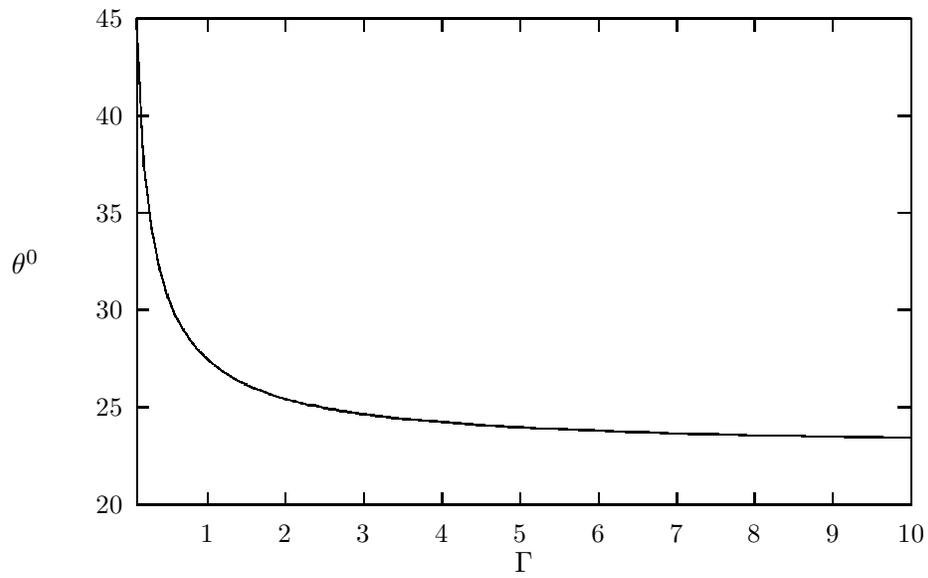}
\caption{\small{Behaviour of the angle of collimation for 
outgoing particles (corotating and counter rotating) which increase their 
energy with respect to infinity, under the influence of a pressure 
gradient. $\Gamma_{i}=0.1$, $\theta_{i}=\pi/4$.}}\label{coll2}
\end{center}
\end{figure}

The explicit form of the variations 
 $(d\Gamma/d\tau)$ and $(dL/d\tau)$ of the orbital parameters, depends on the nature of the perturbation we consider. As shown in 
Paper I,  in the presence of a local 
electromagnetic field or of a pressure gradient, which are the two perturbations we are going to deal with, relations exist which 
link the coordinate $\theta$ to the energy parameter $\Gamma$. 
In case of energy gain ($\frac{d\Gamma}{d\tau}>0$), such relations 
give rise to collimation laws, which we now recall:
\begin{itemize}
\item{ Case of poloidal electromagnetic field (see section 3.1):
\begin{itemize}
\item{cospiralling $\theta=\, const.$ orbits:
\1
\sin{\theta}=\sin{\theta_{i}}\left[\frac{\sqrt{1+1/\Gamma}+1+1/(2\Gamma)}{\sqrt{1+1/\Gamma_{i}}+1
+1/(2\Gamma_{i})}\right]^{1/4}\label{teta1}
\2}
\item{counterspiralling $\theta=\, const.$ orbits:
\1
\sin{\theta}=\sin{\theta_{i}}\left[\frac{\Gamma_{i}(\sqrt{\Gamma_{i}^{2}+\Gamma_{i}}+1/2+\Gamma_{
i})}{\Gamma(\sqrt{\Gamma^{2}+\Gamma}+1/2+\Gamma)}\right]^{1/4}\label{teta2}
\2}
\end{itemize}}
\item{Case of pressure gradients (see section 3.2):
\begin{itemize}
\item{Either case of $\theta=\,const.$ orbits:
\1
\sin{\theta}=\sin{\theta_{i}}\left[\frac{\Gamma_{i}(\Gamma+1)}{\Gamma(\Gamma_{i}+1)}\right]^{1/4}\label{tetaf}
\2}
\end{itemize}}
\end{itemize}
where $\theta_{i}$ and $\Gamma_{i}$ are the initial values. 
In figure \ref{coll1} and \ref{coll2} we show the collimating bahaviour of (\ref{teta1}), (\ref{teta2}) and (\ref{tetaf}) 
under the influence of the corresponding external perturbation which causes an increase of  particle's energy. We have chosen as initial values, 
$\Gamma_{i}=0.1$ at $\theta_{i}=\pi/4$.

Distinction has been made between co-rotating (with the metric source) and 
counter-rotating orbits, collimation being stronger for the latter
\footnote{Curves of figure 2 correct figure 4 of Paper I where they have 
been erroneously crossed. In that figure the two curves had to be considered 
independently and not to be compared.}.
Let us underline the physical meaning of these relations: 
geometry induced collimation, in the presence of an electromagnetic field 
or a pressure gradient, occurs mainly when particles, on initially vortical 
geodesics, increase their energy with respect to infinity. However this is 
not the same as saying that they are locally accelerating, as we are going 
to see.

\section{Test particle acceleration}
Relations (\ref{teta1}) - (\ref{teta2}) - (\ref{tetaf}) pressupose a 
knowledge of how $\Gamma$ varies along the perturbed geodesic under the two kinds of external perturbations  we are here considering. In this way, besides describing the collimation of the vortical geodesics, we will also deduce the behaviour of the local
Lorentz factor $\gamma$ of the particles which leave the neighbourhood of the 
rotating black hole. 

Let us introduce a field of local static observers defined by a four-velocity:
\begin{equation}
\tilde{u} = (-g_{00})^{-1/2} \partial_{t}=e^{\psi}\partial_{t}
\end{equation}
and let $\tilde{k}$ be the particle four-velocity.
The relation between the total particle energy $E$, as measured at 
asymptotic distances, and the fundamental astrophysical quantity 
$\gamma$ locally measured by the observer $\tilde{u}$, is then given by:
\1
\gamma=-(\tilde{u}|\tilde{k})=-e^{\psi}(\partial_{t}|\tilde{k})=e^{\psi}E
\2
that is:
\1
\gamma=\sqrt{\frac{\Gamma+1}{1-\frac{2Mr}{r^{2}+a^{2}\cos^{2}{\theta}}}}
\label{lorentz}
\2
where $M$, as stated, is the mass of the black hole.

For the sake of comparison, let us first consider the behaviour of 
$\gamma$  for a particle in strictly geodesic motion, namely in the 
absence of external perturbations.

\begin{figure}[h]
\begin{center}
% GNUPLOT: LaTeX picture
\setlength{\unitlength}{0.240900pt}
\ifx\plotpoint\undefined\newsavebox{\plotpoint}\fi
\sbox{\plotpoint}{\rule[-0.200pt]{0.400pt}{0.400pt}}%
\begin{picture}(1500,900)(0,0)
\font\gnuplot=cmr10 at 10pt
\gnuplot
\sbox{\plotpoint}{\rule[-0.200pt]{0.400pt}{0.400pt}}%
\put(220.0,113.0){\rule[-0.200pt]{292.934pt}{0.400pt}}
\put(220.0,113.0){\rule[-0.200pt]{4.818pt}{0.400pt}}
\put(198,113){\makebox(0,0)[r]{0}}
\put(1416.0,113.0){\rule[-0.200pt]{4.818pt}{0.400pt}}
\put(220.0,222.0){\rule[-0.200pt]{4.818pt}{0.400pt}}
\put(198,222){\makebox(0,0)[r]{2}}
\put(1416.0,222.0){\rule[-0.200pt]{4.818pt}{0.400pt}}
\put(220.0,331.0){\rule[-0.200pt]{4.818pt}{0.400pt}}
\put(198,331){\makebox(0,0)[r]{4}}
\put(1416.0,331.0){\rule[-0.200pt]{4.818pt}{0.400pt}}
\put(220.0,440.0){\rule[-0.200pt]{4.818pt}{0.400pt}}
\put(198,440){\makebox(0,0)[r]{6}}
\put(1416.0,440.0){\rule[-0.200pt]{4.818pt}{0.400pt}}
\put(220.0,550.0){\rule[-0.200pt]{4.818pt}{0.400pt}}
\put(198,550){\makebox(0,0)[r]{8}}
\put(1416.0,550.0){\rule[-0.200pt]{4.818pt}{0.400pt}}
\put(220.0,659.0){\rule[-0.200pt]{4.818pt}{0.400pt}}
\put(198,659){\makebox(0,0)[r]{10}}
\put(1416.0,659.0){\rule[-0.200pt]{4.818pt}{0.400pt}}
\put(220.0,768.0){\rule[-0.200pt]{4.818pt}{0.400pt}}
\put(198,768){\makebox(0,0)[r]{12}}
\put(1416.0,768.0){\rule[-0.200pt]{4.818pt}{0.400pt}}
\put(220.0,877.0){\rule[-0.200pt]{4.818pt}{0.400pt}}
\put(198,877){\makebox(0,0)[r]{14}}
\put(1416.0,877.0){\rule[-0.200pt]{4.818pt}{0.400pt}}
\put(246.0,113.0){\rule[-0.200pt]{0.400pt}{4.818pt}}
\put(246,68){\makebox(0,0){1}}
\put(246.0,857.0){\rule[-0.200pt]{0.400pt}{4.818pt}}
\put(379.0,113.0){\rule[-0.200pt]{0.400pt}{4.818pt}}
\put(379,68){\makebox(0,0){2}}
\put(379.0,857.0){\rule[-0.200pt]{0.400pt}{4.818pt}}
\put(511.0,113.0){\rule[-0.200pt]{0.400pt}{4.818pt}}
\put(511,68){\makebox(0,0){3}}
\put(511.0,857.0){\rule[-0.200pt]{0.400pt}{4.818pt}}
\put(643.0,113.0){\rule[-0.200pt]{0.400pt}{4.818pt}}
\put(643,68){\makebox(0,0){4}}
\put(643.0,857.0){\rule[-0.200pt]{0.400pt}{4.818pt}}
\put(775.0,113.0){\rule[-0.200pt]{0.400pt}{4.818pt}}
\put(775,68){\makebox(0,0){5}}
\put(775.0,857.0){\rule[-0.200pt]{0.400pt}{4.818pt}}
\put(907.0,113.0){\rule[-0.200pt]{0.400pt}{4.818pt}}
\put(907,68){\makebox(0,0){6}}
\put(907.0,857.0){\rule[-0.200pt]{0.400pt}{4.818pt}}
\put(1039.0,113.0){\rule[-0.200pt]{0.400pt}{4.818pt}}
\put(1039,68){\makebox(0,0){7}}
\put(1039.0,857.0){\rule[-0.200pt]{0.400pt}{4.818pt}}
\put(1172.0,113.0){\rule[-0.200pt]{0.400pt}{4.818pt}}
\put(1172,68){\makebox(0,0){8}}
\put(1172.0,857.0){\rule[-0.200pt]{0.400pt}{4.818pt}}
\put(1304.0,113.0){\rule[-0.200pt]{0.400pt}{4.818pt}}
\put(1304,68){\makebox(0,0){9}}
\put(1304.0,857.0){\rule[-0.200pt]{0.400pt}{4.818pt}}
\put(1436.0,113.0){\rule[-0.200pt]{0.400pt}{4.818pt}}
\put(1436,68){\makebox(0,0){10}}
\put(1436.0,857.0){\rule[-0.200pt]{0.400pt}{4.818pt}}
\put(220.0,113.0){\rule[-0.200pt]{292.934pt}{0.400pt}}
\put(1436.0,113.0){\rule[-0.200pt]{0.400pt}{184.048pt}}
\put(220.0,877.0){\rule[-0.200pt]{292.934pt}{0.400pt}}
\put(45,495){\makebox(0,0){$\gamma$}}
\put(828,23){\makebox(0,0){$\bar{r}$}}
\put(220.0,113.0){\rule[-0.200pt]{0.400pt}{184.048pt}}
\put(220,586){\usebox{\plotpoint}}
\put(269,849){\usebox{\plotpoint}}
\multiput(269.58,803.06)(0.492,-14.153){21}{\rule{0.119pt}{11.067pt}}
\multiput(268.17,826.03)(12.000,-306.031){2}{\rule{0.400pt}{5.533pt}}
\multiput(281.58,505.79)(0.493,-4.263){23}{\rule{0.119pt}{3.423pt}}
\multiput(280.17,512.90)(13.000,-100.895){2}{\rule{0.400pt}{1.712pt}}
\multiput(294.58,403.84)(0.492,-2.392){21}{\rule{0.119pt}{1.967pt}}
\multiput(293.17,407.92)(12.000,-51.918){2}{\rule{0.400pt}{0.983pt}}
\multiput(306.58,350.74)(0.492,-1.487){21}{\rule{0.119pt}{1.267pt}}
\multiput(305.17,353.37)(12.000,-32.371){2}{\rule{0.400pt}{0.633pt}}
\multiput(318.58,317.52)(0.493,-0.933){23}{\rule{0.119pt}{0.838pt}}
\multiput(317.17,319.26)(13.000,-22.260){2}{\rule{0.400pt}{0.419pt}}
\multiput(331.58,294.09)(0.492,-0.755){21}{\rule{0.119pt}{0.700pt}}
\multiput(330.17,295.55)(12.000,-16.547){2}{\rule{0.400pt}{0.350pt}}
\multiput(343.58,276.79)(0.492,-0.539){21}{\rule{0.119pt}{0.533pt}}
\multiput(342.17,277.89)(12.000,-11.893){2}{\rule{0.400pt}{0.267pt}}
\multiput(355.00,264.92)(0.600,-0.491){17}{\rule{0.580pt}{0.118pt}}
\multiput(355.00,265.17)(10.796,-10.000){2}{\rule{0.290pt}{0.400pt}}
\multiput(367.00,254.93)(0.728,-0.489){15}{\rule{0.678pt}{0.118pt}}
\multiput(367.00,255.17)(11.593,-9.000){2}{\rule{0.339pt}{0.400pt}}
\multiput(380.00,245.93)(0.874,-0.485){11}{\rule{0.786pt}{0.117pt}}
\multiput(380.00,246.17)(10.369,-7.000){2}{\rule{0.393pt}{0.400pt}}
\multiput(392.00,238.93)(1.267,-0.477){7}{\rule{1.060pt}{0.115pt}}
\multiput(392.00,239.17)(9.800,-5.000){2}{\rule{0.530pt}{0.400pt}}
\multiput(404.00,233.93)(1.378,-0.477){7}{\rule{1.140pt}{0.115pt}}
\multiput(404.00,234.17)(10.634,-5.000){2}{\rule{0.570pt}{0.400pt}}
\multiput(417.00,228.94)(1.651,-0.468){5}{\rule{1.300pt}{0.113pt}}
\multiput(417.00,229.17)(9.302,-4.000){2}{\rule{0.650pt}{0.400pt}}
\multiput(429.00,224.94)(1.651,-0.468){5}{\rule{1.300pt}{0.113pt}}
\multiput(429.00,225.17)(9.302,-4.000){2}{\rule{0.650pt}{0.400pt}}
\multiput(441.00,220.95)(2.472,-0.447){3}{\rule{1.700pt}{0.108pt}}
\multiput(441.00,221.17)(8.472,-3.000){2}{\rule{0.850pt}{0.400pt}}
\multiput(453.00,217.95)(2.695,-0.447){3}{\rule{1.833pt}{0.108pt}}
\multiput(453.00,218.17)(9.195,-3.000){2}{\rule{0.917pt}{0.400pt}}
\multiput(466.00,214.95)(2.472,-0.447){3}{\rule{1.700pt}{0.108pt}}
\multiput(466.00,215.17)(8.472,-3.000){2}{\rule{0.850pt}{0.400pt}}
\put(478,211.17){\rule{2.500pt}{0.400pt}}
\multiput(478.00,212.17)(6.811,-2.000){2}{\rule{1.250pt}{0.400pt}}
\put(490,209.17){\rule{2.700pt}{0.400pt}}
\multiput(490.00,210.17)(7.396,-2.000){2}{\rule{1.350pt}{0.400pt}}
\put(503,207.17){\rule{2.500pt}{0.400pt}}
\multiput(503.00,208.17)(6.811,-2.000){2}{\rule{1.250pt}{0.400pt}}
\put(515,205.67){\rule{2.891pt}{0.400pt}}
\multiput(515.00,206.17)(6.000,-1.000){2}{\rule{1.445pt}{0.400pt}}
\put(527,204.17){\rule{2.500pt}{0.400pt}}
\multiput(527.00,205.17)(6.811,-2.000){2}{\rule{1.250pt}{0.400pt}}
\put(539,202.67){\rule{3.132pt}{0.400pt}}
\multiput(539.00,203.17)(6.500,-1.000){2}{\rule{1.566pt}{0.400pt}}
\put(552,201.67){\rule{2.891pt}{0.400pt}}
\multiput(552.00,202.17)(6.000,-1.000){2}{\rule{1.445pt}{0.400pt}}
\put(564,200.67){\rule{2.891pt}{0.400pt}}
\multiput(564.00,201.17)(6.000,-1.000){2}{\rule{1.445pt}{0.400pt}}
\put(576,199.17){\rule{2.500pt}{0.400pt}}
\multiput(576.00,200.17)(6.811,-2.000){2}{\rule{1.250pt}{0.400pt}}
\put(588,197.67){\rule{3.132pt}{0.400pt}}
\multiput(588.00,198.17)(6.500,-1.000){2}{\rule{1.566pt}{0.400pt}}
\put(613,196.67){\rule{2.891pt}{0.400pt}}
\multiput(613.00,197.17)(6.000,-1.000){2}{\rule{1.445pt}{0.400pt}}
\put(625,195.67){\rule{3.132pt}{0.400pt}}
\multiput(625.00,196.17)(6.500,-1.000){2}{\rule{1.566pt}{0.400pt}}
\put(638,194.67){\rule{2.891pt}{0.400pt}}
\multiput(638.00,195.17)(6.000,-1.000){2}{\rule{1.445pt}{0.400pt}}
\put(650,193.67){\rule{2.891pt}{0.400pt}}
\multiput(650.00,194.17)(6.000,-1.000){2}{\rule{1.445pt}{0.400pt}}
\put(601.0,198.0){\rule[-0.200pt]{2.891pt}{0.400pt}}
\put(674,192.67){\rule{3.132pt}{0.400pt}}
\multiput(674.00,193.17)(6.500,-1.000){2}{\rule{1.566pt}{0.400pt}}
\put(662.0,194.0){\rule[-0.200pt]{2.891pt}{0.400pt}}
\put(699,191.67){\rule{2.891pt}{0.400pt}}
\multiput(699.00,192.17)(6.000,-1.000){2}{\rule{1.445pt}{0.400pt}}
\put(711,190.67){\rule{3.132pt}{0.400pt}}
\multiput(711.00,191.17)(6.500,-1.000){2}{\rule{1.566pt}{0.400pt}}
\put(687.0,193.0){\rule[-0.200pt]{2.891pt}{0.400pt}}
\put(736,189.67){\rule{2.891pt}{0.400pt}}
\multiput(736.00,190.17)(6.000,-1.000){2}{\rule{1.445pt}{0.400pt}}
\put(724.0,191.0){\rule[-0.200pt]{2.891pt}{0.400pt}}
\put(773,188.67){\rule{2.891pt}{0.400pt}}
\multiput(773.00,189.17)(6.000,-1.000){2}{\rule{1.445pt}{0.400pt}}
\put(748.0,190.0){\rule[-0.200pt]{6.022pt}{0.400pt}}
\put(797,187.67){\rule{3.132pt}{0.400pt}}
\multiput(797.00,188.17)(6.500,-1.000){2}{\rule{1.566pt}{0.400pt}}
\put(785.0,189.0){\rule[-0.200pt]{2.891pt}{0.400pt}}
\put(834,186.67){\rule{2.891pt}{0.400pt}}
\multiput(834.00,187.17)(6.000,-1.000){2}{\rule{1.445pt}{0.400pt}}
\put(810.0,188.0){\rule[-0.200pt]{5.782pt}{0.400pt}}
\put(883,185.67){\rule{3.132pt}{0.400pt}}
\multiput(883.00,186.17)(6.500,-1.000){2}{\rule{1.566pt}{0.400pt}}
\put(846.0,187.0){\rule[-0.200pt]{8.913pt}{0.400pt}}
\put(932,184.67){\rule{3.132pt}{0.400pt}}
\multiput(932.00,185.17)(6.500,-1.000){2}{\rule{1.566pt}{0.400pt}}
\put(896.0,186.0){\rule[-0.200pt]{8.672pt}{0.400pt}}
\put(982,183.67){\rule{2.891pt}{0.400pt}}
\multiput(982.00,184.17)(6.000,-1.000){2}{\rule{1.445pt}{0.400pt}}
\put(945.0,185.0){\rule[-0.200pt]{8.913pt}{0.400pt}}
\put(1055,182.67){\rule{3.132pt}{0.400pt}}
\multiput(1055.00,183.17)(6.500,-1.000){2}{\rule{1.566pt}{0.400pt}}
\put(994.0,184.0){\rule[-0.200pt]{14.695pt}{0.400pt}}
\put(1129,181.67){\rule{2.891pt}{0.400pt}}
\multiput(1129.00,182.17)(6.000,-1.000){2}{\rule{1.445pt}{0.400pt}}
\put(1068.0,183.0){\rule[-0.200pt]{14.695pt}{0.400pt}}
\put(1227,180.67){\rule{2.891pt}{0.400pt}}
\multiput(1227.00,181.17)(6.000,-1.000){2}{\rule{1.445pt}{0.400pt}}
\put(1141.0,182.0){\rule[-0.200pt]{20.717pt}{0.400pt}}
\put(1362,179.67){\rule{3.132pt}{0.400pt}}
\multiput(1362.00,180.17)(6.500,-1.000){2}{\rule{1.566pt}{0.400pt}}
\put(1239.0,181.0){\rule[-0.200pt]{29.631pt}{0.400pt}}
\put(1375.0,180.0){\rule[-0.200pt]{14.695pt}{0.400pt}}
\end{picture}
\caption{Lorentz factor for a particle on a plain geodesic motion. 
$\theta=0.1\, rad$, $E=1.1$}\label{geodetic}
\end{center}
\end{figure}
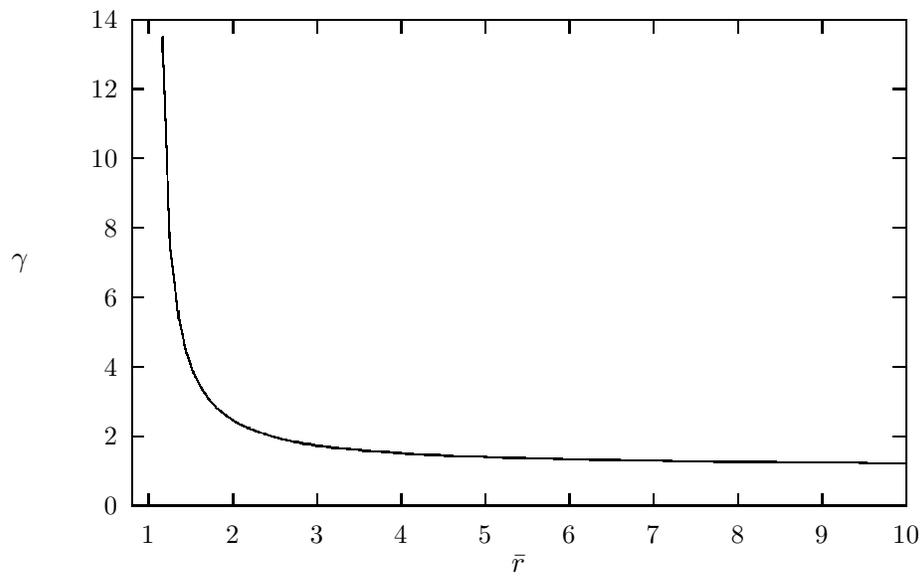

Suppose the particle is moving outwardly on a geodesic with 
$\theta=const=0.1$ rad, arriving at infinity with 
$\gamma_{\infty}=e^{\psi(\infty)}E=E=1.1$, starting from 
$\bar{r_{i}}=r_{i}/M=1.5$ with $\gamma_{i}=1.1\,e^{\psi(\bar{r_{i}},
\theta_{i})}=4.07$ \footnote{
Being $1M_{\odot}=2.2\cdot 10^{3}m$ in geometrized units, 
$\bar{r}_{i}=1.5$ corresponds to a distance from a $10^{8}M_{\odot}$ 
black hole of $r_{i}=\bar{r}_{i}\cdot M=3.3\cdot 10^{11} m\sim 10^{-5}pc$.}.
As we can see from fig \ref{geodetic}, the particle is 
progressively slowed down as seen by a local static observer as it 
moves on its outwardly path. Such a trend reflects the attractive 
character of gravity, according to intuition. 
We have chosen $\bar{a}=a/M=0.9981$. 

In what follows we shall analyse the behaviour of $\gamma$ in a stream of 
particles under the influence of an electromagnetic field and a pressure 
gradient, constrained however by the geodesicity conditions which guarantee 
the simultaneous occurrence of collimation.

\subsection{Effects of the electromagnetic field}
The electromagnetic field we consider arises locally  
from the potential:
\1
A=-\frac{Qr}{\Sigma}(dt-a\sin^{2}{\theta}d\varphi)\label{Carter}
\2
where $Q$ is the total electric charge. 

A charged particle of rest mass $\mu$ and charge $q$ will deviate 
from geodesic motion by the term (neglecting radiation reaction):
\1
\tilde k^{r}\nabla_{r}\tilde k_{i}=\frac{q}{\mu}F_{ij}\tilde k^{j}
\label{Lorentz}
\2
where $F_{ij}=2\partial_{[i}A_{j]}$ and $\tilde{k}$ is, as stated, 
the $4$-velocity of the particle. Since $\tilde k_{0}=-E$, we easly deduce:
\1
-\frac{dE}{d\tau}=\frac{qQ}{\mu\Sigma^{2}}\left[(a^{2}\cos^{2}{\theta}-
r^{2})\frac{dr}{d\tau}+r 
a^{2}\sin{2\theta}\frac{d\theta}{d\tau}\right]\label{dE} 
\2
which tells us how $E$ varies along the perturbed geodesic. 
Since the variation of $E$ is much more sensitive to the variation of the 
coordinate $r$  than to $\theta$, decreasing as $(M/r)^2$ in the first case 
and as $(M/r)^4$ in the latter, we can take as full variation of 
$E$ the following:  
\1
\frac{\partial E}{\partial r}=\frac{qQ}{\mu\Sigma^{2}}(r^{2}-a^{2}
\cos^{2}{\theta})\label{full}
\2
Recalling that $E=\sqrt{\Gamma+1}$ and using normalized quantities, 
relation (\ref{full}) can be written as:
\3
%\frac{1}{M}\frac{\partial E}{\partial\bar{r}}&=&\frac{1}{M}\frac{\partial 
%\sqrt{\Gamma+1}}{\partial\bar{r}}=\frac{qQ}{\mu}\frac{\bar{r}^{2}-
%\bar{a}^{2}\cos^{2}{\theta}}{\bar{\Sigma}^{2}}\frac{1}{M^{2}}\nonumber\\
\frac{\partial \Gamma}{\partial\bar{r}}=2\bar{\cal{C}}\frac{\bar{r}^{2}-
\bar{a}^{2}\cos^{2}{\theta}}{(\bar{r}^{2}+\bar{a}^{2}\cos^{2}{\theta})^{2}}
\sqrt{\Gamma+1}\label{Gam}
\4
where we have put $\bar{C}=\frac{qQ}{\mu M}$.

This is the partial differential equation we were looking for. 
It can be coupled to the laws of collimation (\ref{teta1}) and (\ref{teta2}), 
providing a system in the unknowns $\Gamma(\bar{r})$ and 
$\theta(\Gamma(\bar{r}))$. 
This allows us to reduce (\ref{Gam}) from the form 
$\partial_{\bar{r}}\Gamma=f(\Gamma,\bar{r},\bar{a}\cos{\theta})$ to the form 
$\partial_{\bar{r}}\Gamma=g(\Gamma,\bar{r})$, exploiting the laws of 
collimation we have already found. The new partial differential equation can 
be numerically solved for $\Gamma(\bar{r})$, telling us how $\Gamma$ 
varies along the perturbed geodesic as a function of distance. 
On the other hand, knowledge of $\Gamma(\bar{r})$ allows us to find the 
value of $\theta$ as a function of $r$. Getting the Lorentz factor 
$\gamma=\gamma(\bar{r})$ would be the last step, through 
relation (\ref{lorentz}) with the laws of collimation to be used again.

All seems to be quite smooth, except for the presence of 
parameter $\bar{C}$: in Paper I it was shown that having a magnetic 
field $B\sim 10^{-3} G$ at the distance of $1 pc$ from a $10^{8}M_{\odot}$ 
black hole along the axis, implies $Q\sim 2.3\cdot 10^{11}m$ in 
geometrized units, so that $Q/M\sim 1$. Consequently, $\bar{C}$ is mainly 
the ratio $q/\mu$, which depends critically on the ionization degree.

Since we are interested in the bulk motion of the material, rather than 
the motion of individual particles, we can accept the guiding centre 
approximation and look at our particle as a small cloud of hydrogen, say, 
whose specific charge we need to estimate. 
This requires taking into account the effects of 
photoionization due to the radiation field of the central source, 
the degree of recombination, the optical depth within the cloudlet itself 
and the gravitational redshift of the ionizing radiation. But first of all, 
we ought to know what kind of conditions on  ratio $q/\mu$ comes from 
the geodesicity condition, so essential to our discussion. As stated in 
Paper I,  the geodesic character of the motion can be approximately 
saved if orbital parameters vary slowly in time, geodesicity being 
better preserved where gravity dominates. This led the authors to evaluate 
the changes of energy of an orbit that deviates from a geodesic under 
the effects of an electromagnetic field, written in terms of coordinates 
proper 
values variations (relations (41) of Paper I) as:
\3
\delta E|_{\theta=cost}&=&\frac{\partial E}{\partial r}\delta r\nonumber\\
&=&\frac{q}{\mu}\frac{Q}{M}M\frac{r^{2}-
a^{2}\cos^{2}{\theta}}{(r^{2}+a^{2}\cos^{2}{\theta})^{2}}\delta r\nonumber\\
&=&\frac{q}{\mu}\frac{Q}{M}\left(\frac{M}{r}\right)^{2}\left(
\frac{\Delta}{\Sigma}\right)^{1/2}\frac{1-
(a^{2}/r^{2})\cos^{2}{\theta}}{[1+(a^{2}/r^{2})\cos^{2}{\theta}]^{2}}\delta(\frac{l_{r}}{M})\label{del1}
\4
where $\delta l_r=(g_{rr})^{1/2}\delta r$ is the proper-radial length 
as it would be measured by the 
particle itself. Since a variation of $l_r$ goes on with dynamical time 
while a variation 
of the parameter $E$ goes with the perturbation time, the geodesicity 
conditions require that $\delta E<\delta l_r$.

More precisely, what we ask is that the timescale of a significant 
variation of the physical parameters, $\tau_{var}$, be longer than the 
dynamical time associated with a geodesic trajectory, $\tau_{dyn}$.
Just for an order of magnitude estimate, let us take, as a significant 
dynamical time, the proper time a particle takes to reach the $r=0$ disk 
from the outer horizon on a parabolic trajectory, as was shown in Paper I:
\1
\tau_{dyn}\sim \frac{2M}{3}\left[1+\sqrt{1-\left(\frac{a}{M}\right)^{2}}
\right]\sim \frac{2M}{3}
\2
On the other hand:
\1
\tau_{var}\sim \left(E/\frac{dE}{d\tau}\right) \sim \frac{E}{ \frac{qQ}
{\mu\Sigma^{2}}(r^{2}-a^{2}\cos^{2}{\theta})\frac{dr}{d\tau}}
\2
If we now recover from (A1) of Paper I the $r$ component of the four-vector 
$\tilde{k}$ tangent to a $\theta=0$ geodesic in which we choose, 
for simplicity, $\Gamma\simeq 1$:
\1
k^{r}=\left(\frac{r^{2}+a^{2}+2Mr}{r^{2}+a^{2}}\right)^{1/2}
\2
we get:
\1
\tau_{var}\sim \frac{\sqrt{2}(r^2+a^2)^{5/2}}{\frac{q}{\mu}Q(r^{2}-a^{2})
(r^{2}+a^{2}+2Mr)^{1/2}}
\2
Therefore the geodesicity condition translated into timescale language 
requires that:
\1
\frac{q}{\mu}<\frac{3}{2}\sqrt{2}\frac{(\bar{r}^{2}+\bar{a}^{2})^{5/2}}
{(\bar{r}^{2}-\bar{a}^{2})(\bar{r}^{2}+\bar{a}^{2}+2\bar{r})^{1/2}}
\equiv \bar{C}_{ms}\label{cond1}
\2

This is an explicit, though crude, expression of the geodesicity condition 
in the presence of the particular elettromagnetic field we have considered.
It can be either considered as  a condition on the ratio $q/\mu$ if we fix 
the distance $r$ from the source, or  as  a condition on the distance scale 
if we can fix the ionization. Taking as mentioned $Q/M\sim 1$ and 
$a/M\sim 0.9981$, (\ref{cond1}) tells us  that  $q/\mu$ must be of the 
order $\leq 10$ at short distances from the centre ($r=2M$), while it can 
increase outwardly up to $q/\mu\leq 10^{10}$ at $r\sim 10^{5}M\sim 1pc$.

The astrophysical implications of these requirements have been discussed in 
Paper I, therefore we shall here derive the behaviour of $\gamma(r)$ 
with the assumption that 
the conditions on $q/\mu$ are satisfied\footnote{Since the order of 
magnitude of the ratio $q/\mu$ for an electron is, in geometrized units, 
$q/\mu\sim 10^{21}$, we can deduce that for the godesicity condition to be 
satisfied, we ought to deal with cloudlets which are almost neutral near the 
horizon, and increase their ionization as we move outwards. 
Evidently, we shall refer to  $q/\mu$ as the average degree of ionization in 
a volume element, rather than  to the specific charge carried by a single particle.}.
That is, we are now in a position to integrate equation  (\ref{Gam}) 
and get the behaviour of $\gamma(\bar{r})$ through relation (\ref{lorentz}), 
provided we take  $\bar{\cal {C}}$ values in accordance with condition 
(\ref{cond1}).

To a first approach, we will fix $\bar{\cal C}$ to some constant (and low) 
value, so as to reproduce the behaviour  of an approximately neutral plasma.
Secondly, we will consider $\bar{\cal C}$ linearly rising from the centre 
towards the outer regions, so as to reproduce the behaviour of increasingly 
ionized plasma, as it could be under the effect of a ionizing flux of radiation. 
That such a trend satisfies condition (\ref{cond1}) can be appreciated in 
figure \ref{superposition}, where it is compared with the plot of 
$\bar{C}_{ms}$.

Figures (6)   to (9)  contain the results of the first 
approach. They show the $\gamma(\bar{r})$ profiles
 obtained through the mechanism just outlined, where we have chosen  
$\bar{\cal C}=2.5,5,10$ on a distance scale ranging from $\bar{r_i}=2.3$ 
to $\bar{r}=100$, that is to say from almost outside the event horizon 
to $r\sim 10^{-3}$ pc from the centre. Each figure shows two curves, one 
for co-spiralling orbits (solid line), and the other for counter-spiralling 
ones (dashed line).

Figures (10) and (11)  contain the results of the second approach. 
They show the $\gamma$ profiles obtained with linearly rising 
$\bar{\cal C}$ values, that is to say with $\bar{\cal C}\sim\beta\bar{r}$, 
with
$\beta=0.8$ and $\beta=1.1$. The distance scale has been enlarged up to
the first parsec from the central source. Only one curve has been drawn
in this case, because the profiles for co-rotating and counter-rotating
orbits superimpose at such distances.

The following aspects can be underlined:
\begin{itemize}
\item we are facing an acceleration mechanism, which increases  both $\Gamma$
and $\gamma$ up to asymptotic adjustments in dependence of  parameter
$\bar{\cal C}$. We typically find $\gamma\sim 2\div 4$ at $10^{-3}pc$ from
the centre, and $\gamma\sim 10$ at the distance of $1pc$. For constant
values of $\bar{\cal C}$, slightly higher Lorentz factors are reached on
co-rotating orbits with respect to counter-rotating ones; the difference
disappears on larger scales when $\bar{\cal C}$ is supposed to rise linearly.
\item we can appreciate the distinction between parameter $\Gamma$,
which is linked to  energy  $E$ and which is always raising, and the 
Lorentz factor $\gamma$, which is the quantity measured by the local observer.
Contrary to $\Gamma$, $\gamma$ shows a rapid decay within the first
$r\sim 10M$ from the centre (see figure \ref{C25i}), where the effects of 
gravity are supposed to overwhelm those of the electromagnetic field. As a 
matter of fact, $\gamma$ profiles in this inner region can be almost 
superimposed on those we found for the geodesic motion, proving the
stiffness of geodesic orbits very near the central black hole.
\item In our calculations we have chosen an initial value of $\Gamma_i=0.1$ 
at $\bar{r_i}=2.3$ and $\theta_i=\pi/4$, meaning an  initial value of 
$\gamma_i\sim 2.3$. What can power particles to such a relatively high 
value of the Lorentz factor just outside the event horizon? The question is
still open: some version of the Penrose mechanism has been proposed
(Reva K. 1995, \cite{e}).
\item we should pay attention to the fact that the mechanism we have just
explored coexists with radiation pressure effects. As we have mentioned in
the introduction, inverse Compton interaction of the plasma in the jet with 
the radiation field produces a net deceleration, which always arises when
the Lorentz factor of the bulk motion  exceeds  an equilibrium regime given 
by $\gamma_{eq}\sim 4$, depending on the radiation field distribution one 
adopts. Though in this Paper we have not considered radiation
field as a possible physical perturbation, we can argue, looking at our 
$\gamma$ profiles, that radiation drag will not occur within $10^{-3}pc$ 
from the centre, where $\gamma$ does not exceed the value of $5$. Far from 
the black hole, at $1pc$ from it, we find $\gamma\sim 10$, so there might 
be a perceivable effect of radiation drag tending to lower $\gamma$ to its 
equilibrium value. But far from the black hole, according to 
Sikora et al. (1996), such an effect decreases in importance.
\end{itemize}
\subsection{Effects of the pressure gradient}
Let us now consider the effects on the geodesic motion of the perturbation 
represented by pressure gradients, which probably characterize the inner 
parts of an active galactic nucleus:
\3
p&\sim&\rho^{\alpha}\label{alfa} \\
\rho&\sim&r^{-n}\hspace{2cm}\label{beta}
\4
with $1<n<3$, as suggested by several polytropic models.

These pressures and densities are assumed to be those of a perfect fluid, 
whose elements  approximate the behaviour of the emerging particles.
The relativistic Euler equation leads to a  variation of  the energy of each 
fluid element along the perturbed trajectory, given by:
\1
\frac{dE}{d\tau}=-\frac{E}{p+\rho}\frac{dp}{d\tau}\label{dE2}
\2
For the same reasons we appealed to in the  electromagnetic field 
perturbation case, (\ref{dE2}) transforms into:
\1
 \partial_{r}\Gamma=-\frac{2(\Gamma+1)}{p+\rho}\partial_{r}p\label{Gam2}
\2
which is analogous to equation (\ref{full}). It can be further developed 
if we take $p=A\rho^\alpha$ and $\rho=Br^{-n}$, without specific knowledge 
of the constants $A$ and $B$:
\1
\partial_{\bar{r}}\Gamma=\frac{2n\alpha\bar{\cal D}(\Gamma+1)
\bar{r}^{n(1-\alpha)-1}}{1+{\cal 
D}\bar{r}^{n(1-\alpha)}}\label{Gam3}
\2
where we have put $\bar{\cal D}=\frac{AB^{\alpha-1}}{M^{n(\alpha-1)}}$.

Once again we need to evaluate energy changes on a $\theta=\,const$ 
perturbed geodesic, in order to see what kind of geodesicity condition we 
have to account for. From Paper I we deduce:
\1
\delta E|_{\theta=cost}\leq E\left|\frac{r}{\rho}\frac{\partial p}{\partial 
r}\right|\left(\frac{M}{r}\right)\left(\frac{\Delta}{\Sigma}\right)^{1/2}\delta\left(\frac{l_{r}}{M}\right)\label{dell1}
\2
with the inferred requirement that, at least:
\1
\left|\frac{r}{\rho}\frac{\partial p}{\partial r}\right|<1;\label{cond10}
\2
such a condition is equivalent to the requirement that the sound speed in 
the comoving frame of the fluid is non-relativistic:
\1
v_{s}<\frac{1}{\sqrt{n}}\label{luce}
\2

The last step is now to choose realistic values for $\alpha$ and $n$. 
Theoretical arguments combined to semiempirical estimates of pressure 
based on VLBI measurements (see Begelman 1984, \cite{f}) seem to converge
on $\alpha=2$, $n=2$ 
as the best choice\footnote{$\alpha$ and $n$ are not independent: 
for polytropic gases of the kind we are discussing we know that 
$n=\alpha/(\alpha-1)$.}.
Therefore, the differential equation to integrate is, from (\ref{Gam3}):
\1
\partial_{\bar{r}}\Gamma=\frac{8\bar{\cal D}(\Gamma+1)}{\bar{r}^{3}+
\bar{\cal D}\bar{r}}\label{Gam4}
\2
with the associated geodesicity condition that, from (\ref{luce}), is:
\1
v_s<\frac{1}{\sqrt{2}}\label{luce2}
\2
This latter condition implies a lower limit  to the distance scale where the 
mechanism can operate. In order to see this, let us write down the behaviour 
of $v_s$ upon $r$ as it results if we remember the definition 
$v_{s}^{2}=\partial p/\partial \rho$ with our expressions for pressure and 
density:
\1
v_{s}^{2}\sim \alpha\bar{\cal D}\bar{r}^{-n(\alpha-1)}
\2
and, with our choices for $n$ and $\alpha$:
\1
v_{s}\sim \sqrt{2\bar{\cal D}}\bar{r}^{-1}\label{suono1}
\2
Now, given a value of $\bar{\cal D}$, (\ref{suono1}) 
yields the behaviour of $v_s$, which must satisfy (\ref{luce2}). 
Hence there exists a lower limit in the distance scale, 
given by  $\bar r_{lb}=2\sqrt{\bar{\cal D}}$.

We have numerically solved equation (\ref{Gam4}) using  low values of 
$\bar{\cal D}$. The Lorentz factor $\gamma$
comes from (\ref{lorentz}), with the appropriate law of collimation 
offered by (\ref{tetaf}).
Figures (12), (13) and (14) show the results of our calculations: 
the distance scale extends from a point $\bar{r_{i}}=2\sqrt{\bar{\cal D}}$ 
(in all cases near the event horizon) to  $\sim 10^{-2}$ parsec from the centre.
As we can see, the Lorentz factor shows a maximum at distances an order of 
magnitude farther than the event horizon from the centre, depending on  
parameter $\bar{\cal D}$. At large distances, $\gamma$ tends to decrease, 
adjusting itself to a constant value. The mechanism seems to be most 
efficient in the very central regions of the field.
\section{Comparison with observations}
Some facts make it difficult to speak of an effective comparison with 
observations:
\begin{itemize}
\item what  can be observed and measured, through the analysis of VLBI maps, 
is the proper motion of a radio pattern along the jet. Knowledge of the 
redshift $z$ of the source, combined with a hypothesis on the cosmological 
parameters $q_{0}$ and $H_{0}$, allow us to derive the apparent velocity 
$\beta_{app}$ at which the radio components are seen to be moving. 
Superluminal motion is detected over a  wide distance range, 
starting from $10^{-2}pc$ from the core, up to $10^{2}\div10^{3}pc$ 
far away from it.

The Lorentz factor corresponding to the superluminal motion that is 
usually observed with the above mentioned method is assumed to coincide 
with the Lorentz factor of the plasma bulk motion along the jet, 
responsible for the Doppler boosting of the radiation.
\item in such a manner, statistical analysis carried out on a sample of 
sources, for which superluminal motion has been detected, yield the following 
mean values of the Lorentz factor (taken from data of Ghisellini et al., 1993, \cite{g}) :
\begin{table}[h]
\begin{center}
\begin{tabular}{|l|c|}\hline
Sources & $\gamma$ \\\hline
BL Lac & $10.47\pm1.35$  \\\hline
CDQs & $16.98\pm1.25$  \\\hline
LDQs & $14.45\pm1.38$\\\hline
\end{tabular}
\end{center}
\caption{\small{Mean values of the Lorentz factor for the sample of 
sources in the paper of Ghisellini et al. CDQs and  LDQs refer 
respectively to core dominated quasars and lobe dominated quasars.}}
\end{table}
\newline What these data do not tell us is the distance from the source 
where the Lorentz factor has been measured, while our $\gamma$ values 
refer to distances within the first parsec from the centre.
\item our $\gamma$ profiles have been found in the single-particle approach 
for the case of the electromagnetic field perturbation, and in the 
hydrodynamical approach for the case of the pressure gradient perturbation. 
This guiding centre approximation is the first step towards a full 
magnetohydrodynamical treatment.
\item Lorentz factors $\gamma$ derived from observations are found within 
the framework of special relativity, assuming relativistic motion of the 
outgoing plasma along a direction making an angle $\varphi$ with respect to 
the line of sight. Therefore, we have stressed the importance of having 
introduced a local observer, because around him special relativity holds, 
so that what he measures is approximately what we measure from earth, 
apart from negligible problems due to the choice of a particular local 
observer (we found it useful to choose a local static observer). 
\end{itemize}
What is more, a direct comparison with observations is impossible unless 
we take into account the major role of large scale magnetic fields in 
extracting rotational energy from the black hole. Instead, what can be 
reasonably said is that the mechanisms we have proposed serve as primary 
acceleration mechanisms able to inject the plasma into the magnetosphere 
with values of the Lorentz factor $\gamma$ around $5\div10$.

\section{The distance scale}
The most important restrictions to the efficiency of the acceleration 
mechanisms we have just outlined are represented by the geodesicity conditions, 
which turn out to estabilish conditions on the ionization degree, combined 
with conditions on the distance scale where the mechanisms can operate.

There does not seem  to be any upper limit to the distance scale for the 
mechanisms we have developed, while there is a potential lower limit  for 
both types of perturbation we have considered. Let us justify these 
statements starting from the electromagnetic field perturbation.

\begin{figure}[h]
\begin{center}
\input{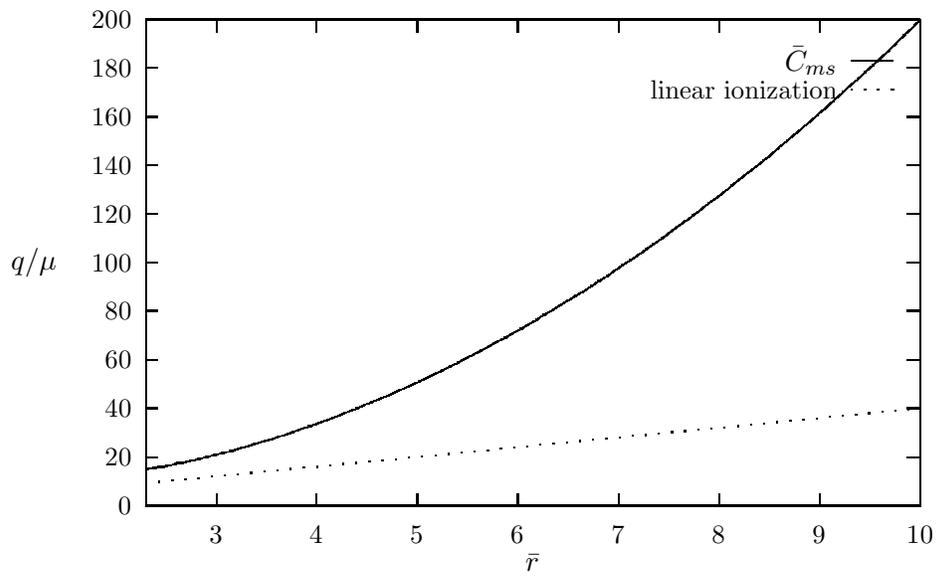}
\caption{\small{The figure shows how  a straight line representing a 
linearly rising ionization ($q/\mu\simeq 4\bar{r}$) is always below the 
curve $\bar{C}_{ms}$ in equation (\ref{cond1}) which  limits the validity 
of geodesicity condition.}}\label{superposition}
\end{center}
\end{figure}

If we come back to figure \ref{superposition} we can infer 
the existence of a lower limit to the distance scale: since the curve 
describes the maximum degree of ionization allowed, a high constant value of 
$q/\mu$ would violate the geodesicity condition at short distances. 
The higher the value of $q/\mu$, the farther the lower limit is from the 
centre. That is the reason why in our calculations we have chosen 
low constant values for  parameter $\bar{\cal C}$. Similarly, a 
linearly rising degree of ionization satisfies the geodesicity condition 
all along the distance scale, as the figure shows.

On the other hand, it is the very nature of the electromagnetic potential  
to prevent the existence of an upper limit to the distance scale. 
In fact, both $A_{t}$ and $A_\varphi$ of  potential (\ref{Carter}) 
vanish at infinity. This suggests that the effects of the electromagnetic 
field would vanish as we moved outwardly\footnote{With a different choice 
of the electromagnetic field, such as that cited by Karas \& Dovciak:
\begin{eqnarray*}
A_{t}&=&\beta a[r\Sigma^{-1}(1+\cos^{2}{\theta})-1]\\
A_{\varphi}&=&\beta\sin^{2}{\theta}[1/2(r^{2}+a^{2})-a^{2}r\Sigma^{-1}
(1+\cos^{2}{\theta})]
\end{eqnarray*}
$\beta\simeq 5\cdot 10^{-8}(B/10^{4}Gauss)(M/10^{8}M_{\odot})\simeq 10^{-8}$, 
the situation would have been different and we would have probably  
found an upper bound too.}  and we have found that the behaviour of $\Gamma$  
confirms this conclusion: $\Gamma$ adjusts itself to a constant value far 
from the centre, as  would be the case of a plain geodesic motion 
($\delta E\rightarrow 0$). Only  the case of a degree of ionization 
intersecting the limiting curve of figure (\ref{superposition}) from 
below would violate the geodesicity condition far from the centre, 
giving rise to an upper limit of the distance scale, but that seems to be 
 quite unphysical.

Now let us turn to the pressure gradient case. We have already shown the 
existence of a lower limit in the distance scale, depending on the 
value of $\bar{\cal D}$: the higher the value of $\bar{\cal D}$, 
the farther the lower limit  from the centre.
No upper limit can be found either: that is physically suggested by the 
fact that pressure gradients vanish at infinity as 
$\partial p/\partial r\sim r^{-5}$ so $v_s$ decreases monotonically in such a way that condition (\ref{luce}) is always satisfied far from the black hole.

In this sense, the acceleration mechanisms we have proposed  need particular 
care near the centre, where we have to hinder the effects of 
extreme physical conditions, while they die  a natural death far from 
the black hole. Therefore, there is no point in extending the 
integration beyond the first parsec from the centre, since neither the 
local  electromagnetic field, nor the pressure gradients have any more 
appreciable effect on the motion at those distances.

\section{Conclusions}
Radio components in jets emerging from active galactic nuclei are found 
to undergo superluminal expansion both near the core, at $10^{-2}pc$ from 
the centre, and far away from it, at $10^{2}-10^{3}pc$. Let us  just 
cite 3C 273, that shows components in superluminal motion at $\sim 50pc$ 
from the core (S.C. Unwin 1989, \cite{i}); 3C 345, studied at $22GHz$, 
which shows superluminal expansion on the parsec scale ($5-25pc$ from the core), 
and 0836+71, which shows superluminal motion at $\sim 220pc$ from the centre.

Therefore, observations suggest that jets are  born with high Lorentz factors. 
In Paper I we indicated  that these structures emerge 
already collimated in the inner region of an AGN, instead of 
being collimated by large scale magnetic fields. In this Paper   
we have tried to complete the picture, testing the 
capability of a properly perturbed gravitational field to produce  
relatively large local Lorentz factors of escaping particles within the 
first parsec from the centre. 

We are now in a position to talk of collimation after acceleration, 
which would occur in the inner region of an AGN, where the effects of the 
black hole space time geometry mix with  those of external 
perturbations, namely a local electromagnetic field and pressure gradients. 
Important restrictions  on the ionization degree of a testing cloudlet 
have been introduced, in order to preserve the geodesic character of the 
motion; specifically, we have found that a degree of  ionization which 
rises linearly with distance from the core is compatible with 
quasi-geodesicity. 

The nature of the interaction between the charged cloudlets and the 
electromagnetic field has been studied  limitedly to the Lorentz forces which arise:   more subtle effects such as Compton losses have been neglected. 
This idealized picture can be justified if we consider the order of 
magnitude of the competing forces, namely  gravoinertial  against Lorentz 
ones. As a consequence, 
the proposed mechanisms show most of their efficiency within the first parsec 
from the centre, producing values of the local Lorentz factor $\gamma$ 
in the range $5\div10$, depending on the parameters introduced. We 
claim that the nozzle appealed to by some hydrodynamical models could 
be a point in the region between $10^{2}\div10^{3}r_{g}$ from the core, 
where we find values of the Lorentz factor between $4\div8$ for the 
two cases of perturbation studied. 

It is to the distance scale to which we refer when we define the above 
developed acceleration mechanisms as primary acceleration mechanisms, 
and we can but confirm the need to  look for different physical 
processes to account for the velocity regime of jets, starting from the 
subparsec region and  ending in the external radio lobes.

\bigskip
{\bf Acknowledgments}
\medskip
Thanks are due to Prof. Mary Evans Prosperi for correcting the English text.

\begin{figure}
\begin{center}
\input{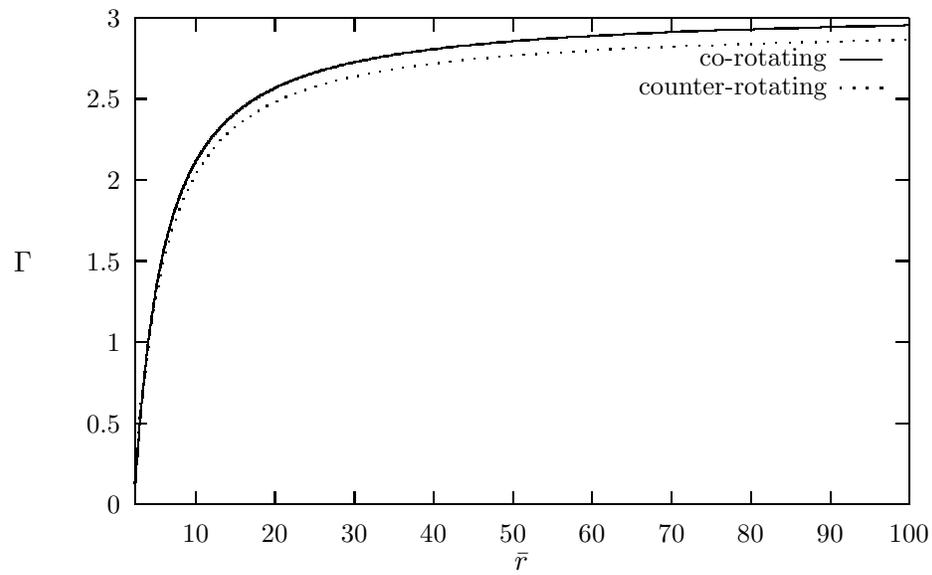}
\caption{\small{Behaviour of $\Gamma$ upon distance for co-rotating 
 and counter-rotating orbits under the effect 
of the electromagnetic field. $\bar{{\cal C}}=2.5$, $\Gamma_{i}=0.1$, 
$\bar{r}_{i}=2.3$.}}\label{C25}
\end{center}
\end{figure}

\begin{figure}
\begin{center}
\input{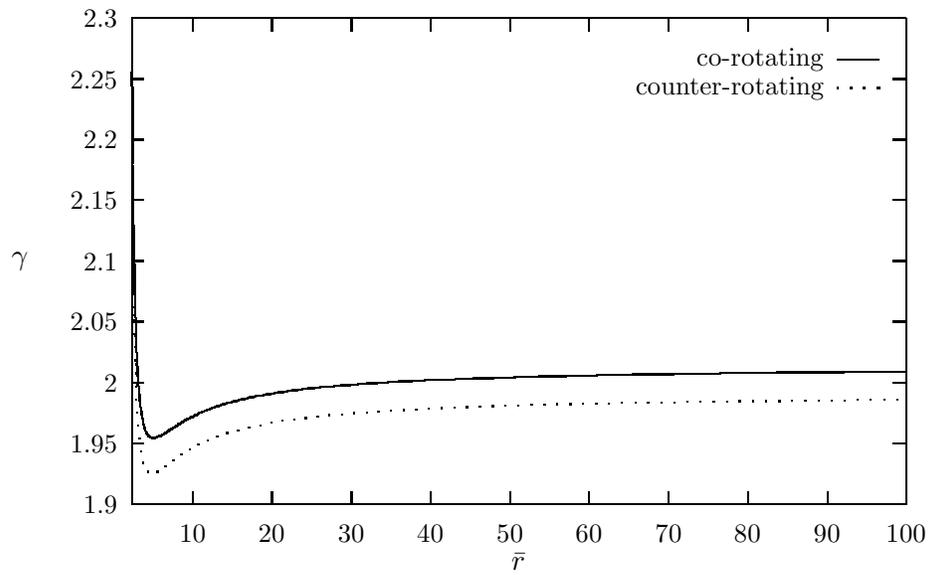}
\caption{\small{Behaviour of the Lorentz factor $\gamma$ 
corresponding to the values of $\Gamma$ represented 
in figure \ref{C25}.}}\label{C25i}
\end{center}
\end{figure}

\begin{figure}
\begin{center}
\input{fig8}
\caption{\small{Behaviour of the Lorentz factor $\gamma$ upon distance 
for co-rotating 
and counter-rotating orbits under the effect 
of the electromagnetic field. $\bar{{\cal C}}=5 $, $\Gamma_{i}=0.1$, 
$\bar{r}_{i}=2.3$.}}\label{C5}
\end{center}
\end{figure}

\begin{figure}
\begin{center}
\input{fig9}
\caption{\small{Behaviour of the Lorentz factor $\gamma$ upon distance 
for co-rotating 
 and counter-rotating  orbits under the 
effect of the electromagnetic field. $\bar{{\cal C}}=10 $, 
$\Gamma_{i}=0.1$, $\bar{r}_{i}=2.3$.}}\label{C10}
\end{center}
\end{figure}

\begin{figure}
\begin{center}
\input{fig10}
\caption{\small{Behaviour of the Lorentz factor $\gamma$  
for co-rotating and counter-rotating orbits with a linearly rising degree 
of ionization: $\bar{\cal C}=0.8\bar{r}$; distance scale enlarged to the 
first parsec from the centre .}}\label{C08}
\end{center}
\end{figure}

\begin{figure}
\begin{center}
\input{fig11}
\caption{\small{Behaviour of the Lorentz factor $\gamma$ for co-rotating 
and counter-rotating orbits with a linearly rising degree of ionization: 
$\bar{\cal C}=1.1\bar{r}$; distance scale enlarged to the first parsec 
from the centre .}}\label{C11}
\end{center}
\end{figure}

\begin{figure}
\begin{center}
\input{fig12}
\caption{\small{Behaviour of the Lorentz factor $\gamma$ 
for co-rotating and counter-rotating orbits under the effect of the 
pressure gradients. $\bar{\cal D}=5$, $\Gamma_{i}=0.1$, 
 $\theta_{i}=\pi/4$.}}\label{p5}
\end{center}
\end{figure}

\begin{figure}
\begin{center}
% GNUPLOT: LaTeX picture
\setlength{\unitlength}{0.240900pt}
\ifx\plotpoint\undefined\newsavebox{\plotpoint}\fi
\sbox{\plotpoint}{\rule[-0.200pt]{0.400pt}{0.400pt}}%
\begin{picture}(1500,900)(0,0)
\font\gnuplot=cmr10 at 10pt
\gnuplot
\sbox{\plotpoint}{\rule[-0.200pt]{0.400pt}{0.400pt}}%
\put(220.0,113.0){\rule[-0.200pt]{0.400pt}{184.048pt}}
\put(220.0,113.0){\rule[-0.200pt]{4.818pt}{0.400pt}}
\put(198,113){\makebox(0,0)[r]{2}}
\put(1416.0,113.0){\rule[-0.200pt]{4.818pt}{0.400pt}}
\put(220.0,222.0){\rule[-0.200pt]{4.818pt}{0.400pt}}
\put(198,222){\makebox(0,0)[r]{3}}
\put(1416.0,222.0){\rule[-0.200pt]{4.818pt}{0.400pt}}
\put(220.0,331.0){\rule[-0.200pt]{4.818pt}{0.400pt}}
\put(198,331){\makebox(0,0)[r]{4}}
\put(1416.0,331.0){\rule[-0.200pt]{4.818pt}{0.400pt}}
\put(220.0,440.0){\rule[-0.200pt]{4.818pt}{0.400pt}}
\put(198,440){\makebox(0,0)[r]{5}}
\put(1416.0,440.0){\rule[-0.200pt]{4.818pt}{0.400pt}}
\put(220.0,550.0){\rule[-0.200pt]{4.818pt}{0.400pt}}
\put(198,550){\makebox(0,0)[r]{6}}
\put(1416.0,550.0){\rule[-0.200pt]{4.818pt}{0.400pt}}
\put(220.0,659.0){\rule[-0.200pt]{4.818pt}{0.400pt}}
\put(198,659){\makebox(0,0)[r]{7}}
\put(1416.0,659.0){\rule[-0.200pt]{4.818pt}{0.400pt}}
\put(220.0,768.0){\rule[-0.200pt]{4.818pt}{0.400pt}}
\put(198,768){\makebox(0,0)[r]{8}}
\put(1416.0,768.0){\rule[-0.200pt]{4.818pt}{0.400pt}}
\put(220.0,877.0){\rule[-0.200pt]{4.818pt}{0.400pt}}
\put(198,877){\makebox(0,0)[r]{9}}
\put(1416.0,877.0){\rule[-0.200pt]{4.818pt}{0.400pt}}
\put(220.0,113.0){\rule[-0.200pt]{0.400pt}{4.818pt}}
\put(220,68){\makebox(0,0){0}}
\put(220.0,857.0){\rule[-0.200pt]{0.400pt}{4.818pt}}
\put(342.0,113.0){\rule[-0.200pt]{0.400pt}{4.818pt}}
\put(342,68){\makebox(0,0){100}}
\put(342.0,857.0){\rule[-0.200pt]{0.400pt}{4.818pt}}
\put(463.0,113.0){\rule[-0.200pt]{0.400pt}{4.818pt}}
\put(463,68){\makebox(0,0){200}}
\put(463.0,857.0){\rule[-0.200pt]{0.400pt}{4.818pt}}
\put(585.0,113.0){\rule[-0.200pt]{0.400pt}{4.818pt}}
\put(585,68){\makebox(0,0){300}}
\put(585.0,857.0){\rule[-0.200pt]{0.400pt}{4.818pt}}
\put(706.0,113.0){\rule[-0.200pt]{0.400pt}{4.818pt}}
\put(706,68){\makebox(0,0){400}}
\put(706.0,857.0){\rule[-0.200pt]{0.400pt}{4.818pt}}
\put(828.0,113.0){\rule[-0.200pt]{0.400pt}{4.818pt}}
\put(828,68){\makebox(0,0){500}}
\put(828.0,857.0){\rule[-0.200pt]{0.400pt}{4.818pt}}
\put(950.0,113.0){\rule[-0.200pt]{0.400pt}{4.818pt}}
\put(950,68){\makebox(0,0){600}}
\put(950.0,857.0){\rule[-0.200pt]{0.400pt}{4.818pt}}
\put(1071.0,113.0){\rule[-0.200pt]{0.400pt}{4.818pt}}
\put(1071,68){\makebox(0,0){700}}
\put(1071.0,857.0){\rule[-0.200pt]{0.400pt}{4.818pt}}
\put(1193.0,113.0){\rule[-0.200pt]{0.400pt}{4.818pt}}
\put(1193,68){\makebox(0,0){800}}
\put(1193.0,857.0){\rule[-0.200pt]{0.400pt}{4.818pt}}
\put(1314.0,113.0){\rule[-0.200pt]{0.400pt}{4.818pt}}
\put(1314,68){\makebox(0,0){900}}
\put(1314.0,857.0){\rule[-0.200pt]{0.400pt}{4.818pt}}
\put(1436.0,113.0){\rule[-0.200pt]{0.400pt}{4.818pt}}
\put(1436,68){\makebox(0,0){1000}}
\put(1436.0,857.0){\rule[-0.200pt]{0.400pt}{4.818pt}}
\put(220.0,113.0){\rule[-0.200pt]{292.934pt}{0.400pt}}
\put(1436.0,113.0){\rule[-0.200pt]{0.400pt}{184.048pt}}
\put(220.0,877.0){\rule[-0.200pt]{292.934pt}{0.400pt}}
\put(45,495){\makebox(0,0){$\gamma$}}
\put(828,23){\makebox(0,0){$\bar{r}$}}
\put(220.0,113.0){\rule[-0.200pt]{0.400pt}{184.048pt}}
\put(223,146){\usebox{\plotpoint}}
\put(222.67,206){\rule{0.400pt}{3.854pt}}
\multiput(222.17,206.00)(1.000,8.000){2}{\rule{0.400pt}{1.927pt}}
\put(223.0,146.0){\rule[-0.200pt]{0.400pt}{14.454pt}}
\put(223.67,351){\rule{0.400pt}{3.614pt}}
\multiput(223.17,351.00)(1.000,7.500){2}{\rule{0.400pt}{1.807pt}}
\put(224.0,222.0){\rule[-0.200pt]{0.400pt}{31.076pt}}
\put(224.67,461){\rule{0.400pt}{2.891pt}}
\multiput(224.17,461.00)(1.000,6.000){2}{\rule{0.400pt}{1.445pt}}
\put(225.0,366.0){\rule[-0.200pt]{0.400pt}{22.885pt}}
\put(225.67,548){\rule{0.400pt}{2.409pt}}
\multiput(225.17,548.00)(1.000,5.000){2}{\rule{0.400pt}{1.204pt}}
\put(226.0,473.0){\rule[-0.200pt]{0.400pt}{18.067pt}}
\put(226.67,615){\rule{0.400pt}{1.927pt}}
\multiput(226.17,615.00)(1.000,4.000){2}{\rule{0.400pt}{0.964pt}}
\put(227.0,558.0){\rule[-0.200pt]{0.400pt}{13.731pt}}
\put(227.67,667){\rule{0.400pt}{1.445pt}}
\multiput(227.17,667.00)(1.000,3.000){2}{\rule{0.400pt}{0.723pt}}
\put(228.0,623.0){\rule[-0.200pt]{0.400pt}{10.600pt}}
\put(228.67,711){\rule{0.400pt}{0.964pt}}
\multiput(228.17,711.00)(1.000,2.000){2}{\rule{0.400pt}{0.482pt}}
\put(229.0,673.0){\rule[-0.200pt]{0.400pt}{9.154pt}}
\put(229.67,740){\rule{0.400pt}{0.964pt}}
\multiput(229.17,740.00)(1.000,2.000){2}{\rule{0.400pt}{0.482pt}}
\put(230.0,715.0){\rule[-0.200pt]{0.400pt}{6.022pt}}
\put(230.67,763){\rule{0.400pt}{0.723pt}}
\multiput(230.17,763.00)(1.000,1.500){2}{\rule{0.400pt}{0.361pt}}
\put(231.0,744.0){\rule[-0.200pt]{0.400pt}{4.577pt}}
\put(231.67,782){\rule{0.400pt}{0.482pt}}
\multiput(231.17,782.00)(1.000,1.000){2}{\rule{0.400pt}{0.241pt}}
\put(232.0,766.0){\rule[-0.200pt]{0.400pt}{3.854pt}}
\put(233,797.67){\rule{0.241pt}{0.400pt}}
\multiput(233.00,797.17)(0.500,1.000){2}{\rule{0.120pt}{0.400pt}}
\put(233.0,784.0){\rule[-0.200pt]{0.400pt}{3.373pt}}
\put(234,808.67){\rule{0.241pt}{0.400pt}}
\multiput(234.00,808.17)(0.500,1.000){2}{\rule{0.120pt}{0.400pt}}
\put(234.0,799.0){\rule[-0.200pt]{0.400pt}{2.409pt}}
\put(235,817.67){\rule{0.241pt}{0.400pt}}
\multiput(235.00,817.17)(0.500,1.000){2}{\rule{0.120pt}{0.400pt}}
\put(235.0,810.0){\rule[-0.200pt]{0.400pt}{1.927pt}}
\put(236,824.67){\rule{0.241pt}{0.400pt}}
\multiput(236.00,824.17)(0.500,1.000){2}{\rule{0.120pt}{0.400pt}}
\put(236.0,819.0){\rule[-0.200pt]{0.400pt}{1.445pt}}
\put(237.0,826.0){\rule[-0.200pt]{0.400pt}{1.445pt}}
\put(237.0,832.0){\usebox{\plotpoint}}
\put(238,836.67){\rule{0.241pt}{0.400pt}}
\multiput(238.00,836.17)(0.500,1.000){2}{\rule{0.120pt}{0.400pt}}
\put(238.0,832.0){\rule[-0.200pt]{0.400pt}{1.204pt}}
\put(239,838){\usebox{\plotpoint}}
\put(239,840.67){\rule{0.241pt}{0.400pt}}
\multiput(239.00,840.17)(0.500,1.000){2}{\rule{0.120pt}{0.400pt}}
\put(239.0,838.0){\rule[-0.200pt]{0.400pt}{0.723pt}}
\put(240,842){\usebox{\plotpoint}}
\put(240.0,842.0){\rule[-0.200pt]{0.400pt}{0.723pt}}
\put(240.0,845.0){\usebox{\plotpoint}}
\put(241.0,845.0){\rule[-0.200pt]{0.400pt}{0.723pt}}
\put(241.0,848.0){\usebox{\plotpoint}}
\put(242,849.67){\rule{0.241pt}{0.400pt}}
\multiput(242.00,849.17)(0.500,1.000){2}{\rule{0.120pt}{0.400pt}}
\put(242.0,848.0){\rule[-0.200pt]{0.400pt}{0.482pt}}
\put(243,851){\usebox{\plotpoint}}
\put(243,851){\usebox{\plotpoint}}
\put(243.0,851.0){\rule[-0.200pt]{0.400pt}{0.482pt}}
\put(243.0,853.0){\usebox{\plotpoint}}
\put(244.0,853.0){\usebox{\plotpoint}}
\put(244.0,854.0){\usebox{\plotpoint}}
\put(245.0,854.0){\rule[-0.200pt]{0.400pt}{0.482pt}}
\put(245.0,856.0){\usebox{\plotpoint}}
\put(246.0,856.0){\usebox{\plotpoint}}
\put(246.0,857.0){\usebox{\plotpoint}}
\put(247.0,857.0){\usebox{\plotpoint}}
\put(247.0,858.0){\usebox{\plotpoint}}
\put(248.0,858.0){\usebox{\plotpoint}}
\put(248.0,859.0){\usebox{\plotpoint}}
\put(249.0,859.0){\usebox{\plotpoint}}
\put(249.0,860.0){\usebox{\plotpoint}}
\put(250.0,860.0){\usebox{\plotpoint}}
\put(250.0,861.0){\rule[-0.200pt]{0.482pt}{0.400pt}}
\put(252.0,861.0){\usebox{\plotpoint}}
\put(252.0,862.0){\rule[-0.200pt]{0.723pt}{0.400pt}}
\put(255.0,862.0){\usebox{\plotpoint}}
\put(255.0,863.0){\rule[-0.200pt]{1.204pt}{0.400pt}}
\put(260.0,863.0){\usebox{\plotpoint}}
\put(260.0,864.0){\rule[-0.200pt]{3.132pt}{0.400pt}}
\put(273.0,863.0){\usebox{\plotpoint}}
\put(273.0,863.0){\rule[-0.200pt]{3.373pt}{0.400pt}}
\put(287.0,862.0){\usebox{\plotpoint}}
\put(287.0,862.0){\rule[-0.200pt]{2.891pt}{0.400pt}}
\put(299.0,861.0){\usebox{\plotpoint}}
\put(299.0,861.0){\rule[-0.200pt]{3.132pt}{0.400pt}}
\put(312.0,860.0){\usebox{\plotpoint}}
\put(312.0,860.0){\rule[-0.200pt]{3.854pt}{0.400pt}}
\put(328.0,859.0){\usebox{\plotpoint}}
\put(328.0,859.0){\rule[-0.200pt]{4.818pt}{0.400pt}}
\put(348.0,858.0){\usebox{\plotpoint}}
\put(348.0,858.0){\rule[-0.200pt]{6.022pt}{0.400pt}}
\put(373.0,857.0){\usebox{\plotpoint}}
\put(373.0,857.0){\rule[-0.200pt]{7.950pt}{0.400pt}}
\put(406.0,856.0){\usebox{\plotpoint}}
\put(406.0,856.0){\rule[-0.200pt]{11.563pt}{0.400pt}}
\put(454.0,855.0){\usebox{\plotpoint}}
\put(454.0,855.0){\rule[-0.200pt]{18.067pt}{0.400pt}}
\put(529.0,854.0){\usebox{\plotpoint}}
\put(529.0,854.0){\rule[-0.200pt]{32.521pt}{0.400pt}}
\put(664.0,853.0){\usebox{\plotpoint}}
\put(664.0,853.0){\rule[-0.200pt]{75.402pt}{0.400pt}}
\put(977.0,852.0){\usebox{\plotpoint}}
\put(977.0,852.0){\rule[-0.200pt]{110.573pt}{0.400pt}}
\end{picture}
\caption{\small{Behaviour of the Lorentz factor $\gamma$ 
for co-rotating and counter-rotating orbits under the effect of the
 pressure gradients. $\bar{\cal D}=10$, $\Gamma_{i}=0.1$, 
$\theta_{i}=\pi/4$.}}\label{p10}
\end{center}
\end{figure}
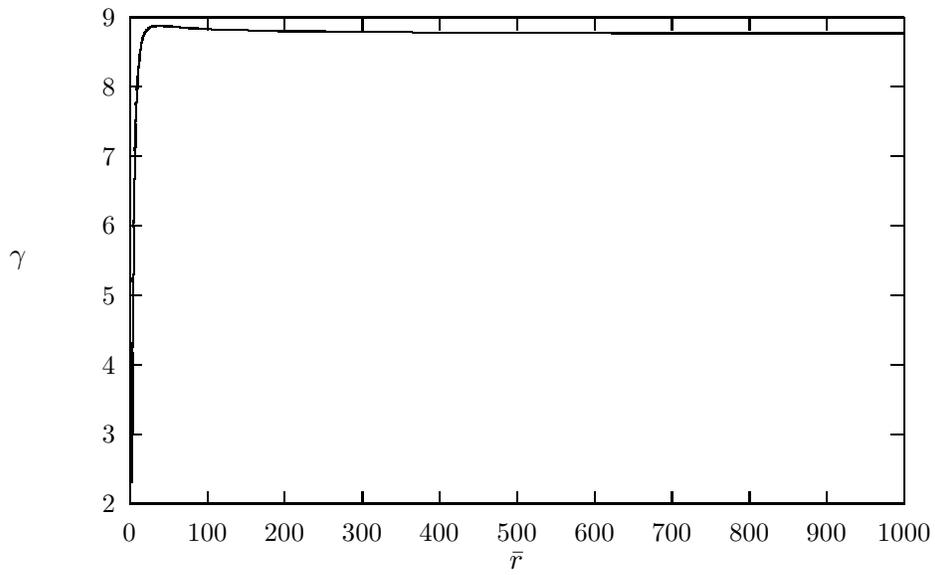

\begin{figure}
\begin{center}
\input{fig14}
\caption{\small{Behaviour of the Lorentz factor $\gamma$ 
for co-rotating and counter-rotating orbits under the effect of the 
pressure gradients. $\bar{\cal D}=15$, $\Gamma_{i}=0.1$, 
$\theta_{i}=\pi/4$.}}\label{p15}
\end{center}
\end{figure}

\end{document}